\documentclass{ifacconf}

\usepackage{graphicx}      
\usepackage{natbib}        
\usepackage{color}
\usepackage{amsfonts} 
\usepackage{amsmath} 
\usepackage{subcaption}
\usepackage{bm}
\usepackage{booktabs}
\usepackage{tikz}
\usepackage{pgfplots}

\usepgfplotslibrary[fillbetween] 
\pgfplotsset{compat=1.15}
\usepgfplotslibrary{external}
\newcommand{\NN}{\mathrm{NN}}
\newcommand{\Normal}[1]{\mathcal{N}\left(#1\right)}
\newcommand{\Ewrt}[2]{\mathbb{E}_{#1}\left[#2\right]}
\newcommand{\KL}[2]{\mathrm{KL}\left(#1||#2\right)}
\newcommand{\mat}[1]{\mathbf{#1}} 

\newif\ifarxiv
\arxivtrue 

\begin{document}
\begin{frontmatter}

\title{Deep State Space Models for Nonlinear System Identification\thanksref{footnoteinfo}} 

\thanks[footnoteinfo]{This research was partially supported by the \emph{Wallenberg AI, Autonomous Systems and Software Program (WASP)} funded by Knut and Alice Wallenberg Foundation and the Swedish Research Council, contracts 2016-06079 and 2019-04956.}

\author[UU]{Daniel Gedon} 
\author[UU]{Niklas Wahlstr\"om} 
\author[UU]{Thomas B. Sch\"on}
\author[Liu]{Lennart Ljung}

\address[UU]{Dept. of Information Technology, Uppsala University, Sweden E-mail: \{daniel.gedon, niklas.wahlstrom, thomas.schon\}@it.uu.se}
\address[Liu]{Div. of Automatic Control, Link\"oping University, Sweden E-mail: lennart.ljung@liu.se}

\begin{abstract}
Deep state space models (SSMs) are an actively researched model class for temporal models developed in the deep learning community which have a close connection to classic SSMs. The use of deep SSMs as a black-box identification model can describe a wide range of dynamics due to the flexibility of deep neural networks. Additionally, the probabilistic nature of the model class allows the uncertainty of the system to be modelled. In this work a deep SSM class and its parameter learning algorithm are explained in an effort to extend the toolbox of nonlinear identification methods with a deep learning based method. Six recent deep SSMs are evaluated in a first unified implementation on nonlinear system identification benchmarks. 
\end{abstract}


\end{frontmatter}

\section{Introduction}

System identification is a well-established area of automatic control, see \cite{astromSystemIdentificationSurvey1971,ljungSystemIdentificationTheory1999}. A wide range of identification methods have been developed for parametric and non-parametric models as well as for grey-box and black-box models. Contrary, the field of machine learning and deep learning has emerged as the new standard in many disciplines to model highly complex systems, see \cite{goodfellow2016DeepLearning}. Deep learning can identify and capture patterns as a black-box model. It has been shown to be useful for high dimensional and nonlinear problems emerging in diverse areas such as image analysis, time series modelling, speech recognition and text classification. This paper provides one step to combine the areas of system identification and deep learning by showing the usefulness of deep SSMs applied to nonlinear system identification. It helps to bridge the gap between the fields and to learn from each others advances.

Nowadays, a wide range of system identification algorithms for parametric models are available. Parametric models such as SSMs can include pre-existing knowledge about the structure of the system and can result in precise identification. For automatic control this is a popular model class and a variety of identification algorithms is available, e.g. \cite{schonSystemIdentificationNonlinear2011}.

In deep learning recent advances in the development of deep SSMs have been made, e.g. \cite{bayerLearningStochasticRecurrent2015,chungRecurrentLatentVariable2015,fraccaroSequentialNeuralModels2016}. The class of deep SSMs has three main advantages. (1) Compared with SSMs it is more flexible due to the use of Neural Networks (NNs). (2) 
In many cases it can be more expressive for temporal data than feedforward NNs because of its recurrent structure including hidden states.
(3) Deep SSMs can capture the output uncertainty. These advantages have been exploited for the generation of handwritten text by \cite{liwickiIAMOnDBOnlineEnglish2005} and speech by \cite{prahalladBlizzardChallenge20132013}. The examples have highly nonlinear dynamics and
require accurate uncertainty quantification to generate new realistic sequences.
Our main contributions are:
\begin{itemize}
    \item Bring the communities of system identification and deep learning closer by rigorously elaborating a deep learning model class and its learning algorithm, while applying it to nonlinear system identification problems. It extends the toolbox of possible identification approaches with a new class of deep learning models. This paper complements the work by \cite{anderssonDeepConvolutionalNetworks2019}, where deterministic NNs are applied to nonlinear system identification.
    \item In system identification there is a clear separation of model structure and parameter estimation. In this paper the same distinction between model structure (Section~\ref{sec2}) and parameter learning (Section~\ref{sec3}) is taken as a future guideline for deep learning.
    \item Six deep SSMs are compared in a unified implementation for nonlinear system identification (Section~\ref{sec4}). The model advantages are highlighted by showing that a maximum likelihood estimate is obtained and additionally the uncertainty is captured which is beneficial in robust control or system analysis.
\end{itemize}
\section{Deep State Space Models for Sequential Data}
\label{sec2}

Sequence modeling is an active topic in deep learning as motivated by the temporal nature of the physical environment. A dynamic model is required to encode the system dynamics. The model is identified from observed input-output pairs $\{(\mat{u}_t, \mat{y}_t)\}_{t=1}^T$ to predicted outputs $\mat{\hat{y}}_t$. An SSM is obtained if the computations are performed via a latent variable $\mat{h}$ that incorporates past information:
\begin{subequations}
\begin{align}
	\mat{h}_t &= f_\theta(\mat{h}_{t-1},\mat{u}_t,\mat{y}_t), \label{eq2:DeepSSM:state}\\
	\mat{\hat y}_t &= g_\theta(\mat{h}_t),\label{eq2:DeepSSM:output}
\end{align}
\end{subequations}
where $\theta$ are unknown parameters. If the functions $f_\theta(\cdot)$ and $g_\theta(\cdot)$ are described by deep mappings such as deep NNs, the resulting model is referred to as a deep SSM.

Another deep learning research direction is that of \emph{generative models} involving generative adversarial networks (GANs) by \cite{goodfellow2014GenerativeAdversarialNets} and Variational Autoencoders (VAEs) by \cite{kingmaAutoEncodingVariationalBayes2013}, which are used to learn representations of the data and generate new instances from the same distribution, such as realistic images. Extending VAEs to sequential models such as in \cite{fraccaroDeepLatentVariable2018} yields the subclass of deep SSM models which are studied in this paper. The building  blocks for these models are Recurrent NNs (RNNs) and VAEs.

\subsection{Recurrent Neural Networks}

RNNs are useful in modeling sequences of variable length. Models with external inputs $\mat{u}_t$ and outputs $\mat{y}_t$ at each time step are considered. RNNs make use of a hidden state $\mat{h}_t=f_\theta(\mat{h}_{t-1},\mat{u}_t)$. The function parameters are learned by unfolding the RNN and using backpropagation through time. The most notable types of RNNs 
for long-term dependencies are Long Short-Term Memory (LSTM) networks by \cite{hochreiterLongShortTermMemory1997} and Gated Recurrent Units (GRUs) by \cite{choPropertiesNeuralMachine2014}, which yield empirically similar results. 
GRUs are used in this paper due to their structural simplicity.

\subsection{Variational Autoencoders}

\begin{figure}[thpb]
    \centering
    \includegraphics[width=0.38\textwidth]{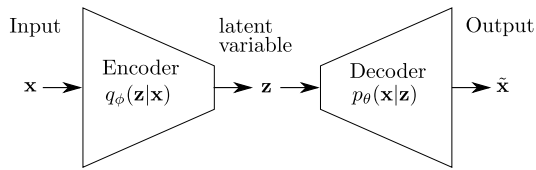}
    \caption{Conceptual idea of the VAE.}
    \label{fig2.2:vae:general}
\end{figure}

A VAE 
embeds a representation of the data distribution of $\mat{x}$ in a low dimensional latent variable $\mat{z}$ via an inference network (encoder). A decoder network uses $\mat{z}$ to generate new data $\mat{\widetilde{x}}$ of approximately the same distribution as $\mat{x}$. 
The conceptual idea of a VAE is visualized in Fig.~\ref{fig2.2:vae:general} and can be viewed as a latent variable model through $\mat{z}$.

Within VAEs it is generally assumed that the data $\mat{x}$ can be modelled by a normal distribution. The decoder is chosen accordingly as $p_\theta(\mat{x}|\mat{z})=\Normal{\mat{x}|\bm{\mu}^{\mathrm{dec}},\bm{\sigma}^{\mathrm{dec}}}$. The parameters for this distribution are given by $[\bm{\mu}^\mathrm{dec}, \bm{\sigma}^\mathrm{dec}]=\NN^\mathrm{dec}_\theta(\mat{z})$ as a deep NN with parameters $\theta$, input $\mat{z}$ and outputs $\bm{\mu}^\mathrm{dec}$ and $\bm{\sigma}^\mathrm{dec}$. The generative model is characterized by the joint distribution $p_\theta(\mat{x},\mat{z}) = p_\theta(\mat{x}| \mat{z})p_\theta(\mat{z})$, where the multivariate normal distribution $p_\theta(\mat{z})=\mathcal{N}(\mat{z}|\bm{\mu}^{\mathrm{prior}},\bm{\sigma}^{\mathrm{prior}})$ is used as prior. The prior parameters are usually chosen as $[\bm{\mu}^{\mathrm{prior}},\bm{\sigma}^{\mathrm{prior}}] = [\mat{0},\mat{I}]$.

For the data embedding in $\mat{z}$, the distribution of interest is the posterior $p(\mat{z}|\mat{x})$ which is intractable in general. It is approximated by a parametric distribution $q_\phi(\mat{z}|\mat{x})=\Normal{\mat{z}|\bm{\mu}^\mathrm{enc},\bm{\sigma}^\mathrm{enc}}$. The distribution parameters are encoded by a deep NN $[\bm{\mu}^\mathrm{enc},\bm{\sigma}^\mathrm{enc}]=\NN^\mathrm{enc}_\phi(\mat{x})$. This network is optimized by variational inference of the variational parameters $\phi$ shared over all data points, \cite{bleiVariationalInferenceReview2017}.

Notably, there exists a connection between the VAE and linear dimension reduction methods such as PCA. In \cite{roweisEMAlgorithmsPCA1998} it is shown that the PCA corresponds to a linear Gaussian model. Specifically, the VAE can be viewed as a nonlinear generalization of the probabilistic PCA. 

\subsection{Combining RNNs and VAEs into deep SSMs}
\begin{figure}[htb!]
	\centering
	\begin{subfigure}[t]{0.25\textwidth}
		\centering
		\includegraphics[width=0.5\textwidth]{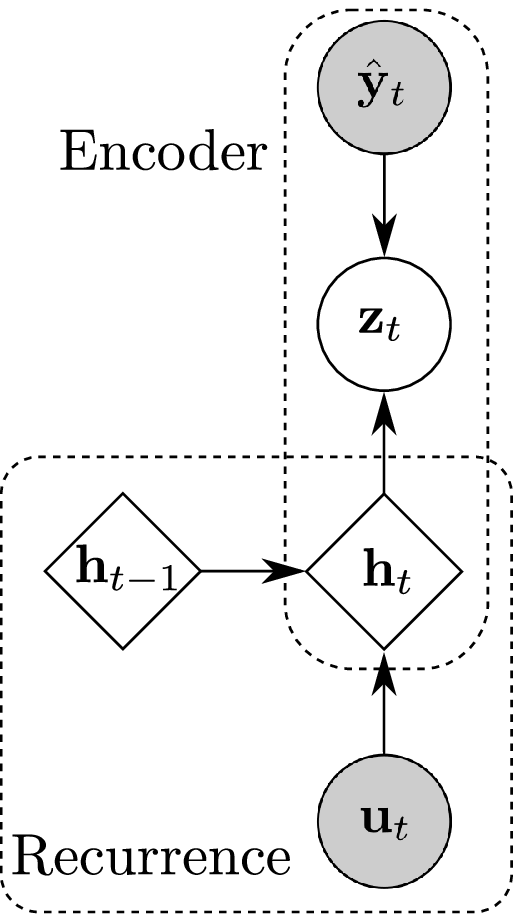}
		\caption{Inference network}
	\end{subfigure}%
	~
	\begin{subfigure}[t]{0.25\textwidth}
		\centering
		\includegraphics[width=0.5\textwidth]{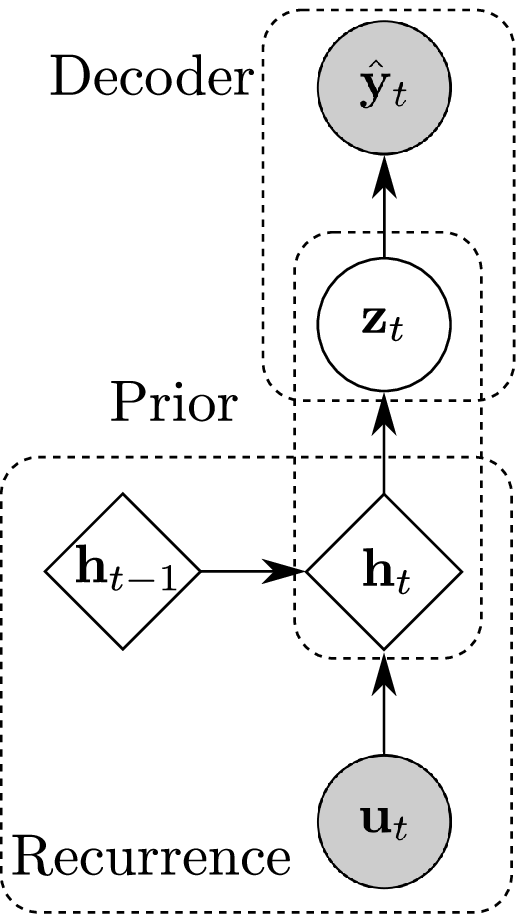}
		\caption{Generative network}
	\end{subfigure}	
	\caption{Graphical model of the VAE-RNN model. \ifarxiv Round blocks indicate probabilistic variables and rectangular blocks deterministic variables. Shaded blocks indicate observed variables.\fi}
	\label{fig2:vaernn:general}
\end{figure}

To obtain a deep SSM we combine RNNs with VAEs, see Fig.~\ref{fig2:vaernn:general} for concrete examples.
The RNN can be viewed as a special case of classic SSMs with Dirac delta functions as state transition distribution $\widetilde{p}(\mat{h}_t|\mat{h}_{t-1})$ compare with \eqref{eq2:DeepSSM:state} or \cite{fraccaroDeepLatentVariable2018}. The VAE can be used to approximate the output distributions of the dynamics from the RNN output, see \eqref{eq2:DeepSSM:output}. A temporal extension of the VAE is required for the studied class of deep SSMs. The parameters of the VAE prior are updated sequentially with the output $\mat{z}_t$ of the RNN as $[\bm{\mu}_t^{\mathrm{prior}},\bm{\sigma}_t^{\mathrm{prior}}]=\NN^\mathrm{prior}_\theta(\mat{z}_{t-1},\mat{u}_t)$. The state transition distribution is given by $p_\theta(\mat{z}_t|\mat{z}_{t-1}, \mat{u}_t) = \mathcal{N}(\mat{z}_t|\bm{\mu}_t^{\mathrm{prior}},\bm{\sigma}_t^{\mathrm{prior}})$. Note that compared with the VAE prior, the parameters $\bm{\mu}_t^{\mathrm{prior}},\bm{\sigma}_t^{\mathrm{prior}}$ are now not static but dependent on previous time steps and describe the recurrence of the model. Similarly the output distribution is given as $p_\theta(\mat{y}_t|\mat{z}_t) = \mathcal{N}(\mat{y}_t|\bm{\mu}_t^{\mathrm{dec}},\bm{\sigma}_t^{\mathrm{dec}})$ with $[\bm{\mu}_t^{\mathrm{dec}},\bm{\sigma}_t^{\mathrm{dec}}]=\NN^\mathrm{dec}_\theta(\mat{z}_t)$. The joint distribution of the deep SSM is
\begin{align}
    p_\theta(\mat{y}_{1:T},\mat{z}_{1:T}|\mat{u}_{1:T},\mat{z}_0)= \prod_{t=1}^T p_\theta(\mat{y}_t|\mat{z}_t)p_\theta(\mat{z}_t|\mat{z}_{t-1},\mat{u}_t).\label{Eq2C:truePosteriorJoint}
\end{align}
Similar to the VAE, this expression describes the generative process. It can be further decomposed with a clear separation between the RNN and the VAE which yields the most simple form within the studied class of deep SSMs, the so-called VAE-RNN from \cite{fraccaroDeepLatentVariable2018}. The model consists of stacking a VAE on top of an RNN as shown in Fig.~\ref{fig2:vaernn:general}. Notice the clear separation between model parameter learning in the inference network with the available data $\{(\mat{u}_t,\mat{y}_t)\}_{t=1}^T$ and the output prediction $\mat{\hat{y}}_t$ in the generative network. The joint true posterior can be factorized according to the graphical model as
\begin{align}
	p_\theta(\mat{y}_{1:T},\mat{z}_{1:T},\mat{h}_{1:T}|\mat{u}_{1:T},\mat{h}_0) &= p_\theta(\mat{y}_{1:T}|\mat{z}_{1:T}) \times  \notag\\ 
	\times p_\theta(\mat{z}_{1:T}|\mat{h}_{1:T})&\widetilde{p}(\mat{h}_{1:T}|\mat{u}_{1:T},\mat{h}_0),\label{eq2:VAERNN:pfactorization}
\end{align}
with prior given by $p_\theta(\mat{z}_t|\mat{h}_t)= \Normal{\mat{z}_t|\bm{\mu}_t^\mathrm{prior},\bm{\sigma}_t^\mathrm{prior}}$ with $[\bm{\mu}_t^\mathrm{prior},\bm{\sigma}_t^\mathrm{prior}]= \NN^\mathrm{prior}_\theta(\mat{h}_t)$ only depending on the recurrent state $\mat{h}_t$.
The approximate posterior can be chosen to mimic the same factorization
\begin{align}
	q_\phi(\mat{z}_{1:T},\mat{h}_{1:T}|\mat{y}_{1:T},\mat{u}_{1:T},\mat{h}_0) &= q_\phi(\mat{z}_{1:T}|\mat{y}_{1:T},\mat{h}_{1:T})\times \notag \\
	\times& \widetilde{p}(\mat{h}_{1:T}|\mat{u}_{1:T},\mat{h}_0).\label{eq2:VAERNN:qfactorization}
\end{align}
In this paper we consider variations in this class of the deep SSM next to the VAE-RNN, specifically:
\begin{itemize}
	\item Variational RNN (VRNN) by \cite{chungRecurrentLatentVariable2015}: Based on VAE-RNN but the recurrence additionally uses the previous latent variable $\mat{z}_{t-1}$ for $p_\theta(\mat{h}_t) = p_\theta(\mat{h}_t|\mat{h}_{t-1},\mat{u}_t,\mat{z}_{t-1})$.
	\item VRNN-I by \cite{chungRecurrentLatentVariable2015}: Same as VRNN but a static prior is used $[\bm{\mu}^\mathrm{prior}, \bm{\sigma}^\mathrm{prior}]=[\mat{0},\mat{I}]$ in every time step.
	\item Stochastic RNN (STORN) by \cite{bayerLearningStochasticRecurrent2015}: Based on the VRNN-I. In the inference network STORN additionally makes use of a forward running RNN with input $\mat{y}_t$, latent variable $\mat{d}_t$ and output $\mat{z}_t$. Hence, $\mat{z}_t$ is characterized by $p_\theta(\mat{z}_t) = \int p_\theta(\mat{z}_t|\mat{d}_t)p_\theta(\mat{d}_t|\mat{d}_{t-1},\mat{y}_t)d\mat{d}_t$.
\end{itemize}
\ifarxiv Graphical models for these extensions are provided in Appendix~\ref{app:A}. \fi
For VRNN and VRNN-I an additional version using Gaussian mixtures as output distribution (VRNN-GMM) is studied. More methods are available in literature, see e.g. \cite{aliasparthgoyalZForcingTrainingStochastic2017,fraccaroSequentialNeuralModels2016}.

\section{Model Parameter Learning}
\label{sec3}

\subsection{Cost Function for the VAE}

The parameter learning method of the deep SSMs is based on the method used for VAEs. The VAE parameters $\theta$ are learned by maximum likelihood $\mathcal{L}(\theta) = \sum_{i=1}^N \log p_\theta(x_i) = \sum_{i=1}^N \mathcal{L}_i(\theta)$ with $N$ data points $\{x_i\}_{i=1}^N$. Performing variational inference with shared parameters for all data results in the following using Jensen's inequality
\begin{align}
	\mathcal{L}_i(\theta) &=\log p_\theta(\mat{x}) = \log \int p_\theta(\mat{x},\mat{z}) d\mat{z} \notag\\
	&= \log \Ewrt{q_\phi(\mat{z}|\mat{x})}{\frac{p_\theta(\mat{x},\mat{z})}{q_\phi(\mat{z}|\mat{x})}} \notag\\
	&\geq \Ewrt{q_\phi(\mat{z}|\mat{x})}{\log\frac{p_\theta(\mat{x},\mat{z})}{q_\phi(\mat{z}|\mat{x})}} = \widetilde{\mathcal{L}}_i(\theta,\phi).\label{eq2B:VAEJensensIneq}
\end{align}
The expression $\widetilde{\mathcal{L}}_i(\theta,\phi)$ is referred to as the evidence lower bound (ELBO) and can be rewritten using the Kullback-Leibler (KL) divergence
\begin{align}
	\widetilde{\mathcal{L}}_i(\theta,\phi) = \Ewrt{q_\phi}{\log p_\theta(\mat{x}|\mat{z})}  -\KL{q_\phi(\mat{z}|\mat{x})}{p_\theta(\mat{z})},\label{eq3:VAEtotalEBO}
\end{align}
where the expectation is w.r.t. $q_\phi(\mat{z}|\mat{x})$. The first term encourages the reconstruction of the data by the decoder. The KL-divergence in the second term is a measure of closeness between the two distributions and can be interpreted as a regularization term. Approximate posteriors $q_\phi(\mat{z}|\mat{x})$ far away from the prior $p_\theta(\mat{z})$ are penalized. The ELBO is then given by $\widetilde{\mathcal{L}}(\theta,\phi)=\sum_{i=1}^N\widetilde{\mathcal{L}}_i(\theta,\phi)$ which is maximized instead of the intractable log-likelihood $\mathcal{L}(\theta)$.

\subsection{Cost Function for Deep SSMs}

A temporal extension of the VAE parameter learning is required for the studied deep SSMs. A similar derivation for the ELBO of the VAE as in \eqref{eq2B:VAEJensensIneq} leads for the generic deep SSM to
\begin{align}
\widetilde{\mathcal{L}}(\theta,\phi) = 
\Ewrt{q_\phi}{\log \frac{p_\theta(\mat{y}_{1:T},\mat{z}_{1:T}|\mat{u}_{1:T},\mat{z}_0)}{q_\phi(\mat{z}_{1:T}|\mat{y}_{1:T},\mat{u}_{1:T},\mat{z}_0)}},
\end{align}
where the expectation is w.r.t. the approximate distribution $q_\phi(\mat{z}_{1:T}|\mat{y}_{1:T},\mat{u}_{1:T},\mat{z}_0)$. The factorization of the true joint posterior distribution from \eqref{Eq2C:truePosteriorJoint} can be applied which yields an ELBO as the sum over all time steps. Note that in this generic scheme $q_\phi(\cdot)$ can be factorized as $\prod_{t=1}^T q_\phi(\mat{z}_t|\mat{z}_{t-1},\mat{y}_{t:T},\mat{u}_{t:T})$, which requires a smoothing step since $\mat{z}_t$ depends on all inputs and outputs for all time steps $t=1,\dots,T$. If there exists a similar factorization for the approximate posterior as in \eqref{eq2:VAERNN:qfactorization}, then an expression similar to \eqref{eq3:VAEtotalEBO} for the VAE in can be obtained.

In the VAE-RNN a solution for parameter learning is obtained by a clear separation between the RNN and the VAE. 
Note that here no smoothing step for the variational distribution is necessary since the states $\mat{z}_{1:T}$ are independent given $\mat{h}_{1:T}$ as can be seen by d-separation in Fig.~\ref{fig2:vaernn:general}. The same factorization as in \eqref{eq2:VAERNN:qfactorization} can be used. The ELBO for the VAE-RNN is written as
\begin{align}
    \widetilde{\mathcal{L}}(\theta,\phi)=
    \Ewrt{q_\phi}{\log \frac{p_\theta(\mat{y}_{1:T},\mat{z}_{1:T},\mat{h}_{1:T}|\mat{u}_{1:T},\mat{h}_0)}{q_\phi(\mat{z}_{1:T},\mat{h}_{1:T}|\mat{y}_{1:T},\mat{u}_{1:T},\mat{h}_0)}},\label{Eq3:TotalELBOfull}
\end{align}
where the expectation is taken w.r.t. the approximate posterior $q_\phi(\mat{z}_{1:T},\mat{h}_{1:T}|\mat{y}_{1:T},\mat{u}_{1:T},\mat{h}_0)$. Applying the posterior factorizations in \eqref{eq2:VAERNN:pfactorization} and \eqref{eq2:VAERNN:qfactorization} to the ELBO in \eqref{Eq3:TotalELBOfull} and taking the expectation w.r.t. $q_\phi(\mat{z}_t|\mat{y}_t,\mat{h}_t)$ yields
\begin{align}
    \widetilde{\mathcal{L}}(\theta,\phi) = \sum_{t=1}^T &\Ewrt{q_\phi}{\log \frac{p_\theta(\mat{y}_t|\mat{z}_t)p_\theta(\mat{z}_t|\mat{h}_t)}{q_\phi(\mat{z}_t|\mat{y}_t,\mat{h}_t)}} \notag\\
        = \sum_{t=1}^T&\Ewrt{q_\phi}{\log p_\theta(\mat{y}_t|\mat{z}_t)} - \notag \\ 
        &\KL{q_\phi(\mat{z}_t|\mat{y}_t,\mat{h}_t)}{p_\theta(\mat{z}_t|\mat{h}_t)},
\end{align}
which is of the same form as the VAE ELBO in \eqref{eq3:VAEtotalEBO}, but with a temporal extension summing over all time steps.
\section{Numerical Experiments}
\label{sec4}

All six models described in Section~\ref{sec2} are evaluated. The model hyperparameters are the dimension of the hidden state $\mat{z}_t$ denoted by $z_\mathrm{dim}$, the dimension of the RNN hidden state $\mat{h}_t$ denoted by $h_\mathrm{dim}$ and the number of layers within the RNN networks $n_\mathrm{layer}$. For STORN the dimension of $\mat{d}_t$ is chosen equal to that of $\mat{h}_t$. The VRNN-GMM uses five Gaussian mixtures in the output distribution. The encoder and decoder are modelled as 3-layer NN and the features of $\mat{y}_t$, $\mat{u}_t$, $\mat{z}_t$ are extracted with 2-layer NNs.

For parameter learning, hyperparameter and model selection, the data is split in training and validation data. A separate test data set is used for evaluating the final performance. The ADAM optimizer with default parameters is used with early stopping and batch normalization, see \cite{kingmaAdamMethodStochastic2017}. The initial learning rate of $10^{-3}$ is decreased if the validation loss plateaus. Note that the optimization parameters are not fine-tuned for any for the experiments whereas the sequence length for training is considered to be a design parameter.

Three experiments are conducted: (1) a linear Gaussian system, (2) the nonlinear Narendra-Li Benchmark from \cite{narendra1996nonlinearbenchmark}, and (3) the Wiener-Hammerstein (WH) process noise benchmark from \cite{schoukensWienerHammersteinBenchmarkProcess2016}. The first two experiments are considered to show the power of deep SSMs for uncertainty quantification with known true uncertainty, while the last experiment serves as a more complex real world example.
The identified models are evaluated in open loop. The initial state is not estimated. The generated output sequences are compared with the true test data output. As performance metric, the root mean squared error (RMSE) is considered, $\sqrt{\frac{1}{T}\sum_{t=1}^{T}(\mat{\hat{y}}_t-\mat{y}_t)^2}$ with $\mat{\hat{y}}_t=\bm{\mu}_t^\mathrm{dec}$ such that a fair comparison with maximum likelihood estimation methods can be made. To quantify the quality of the uncertainty estimate, the negative log-likelihood (NLL) per time step is used, $\frac{1}{T}\sum_{t=1}^{T}-\log \Normal{\mat{y}_t|\bm{\mu}_t^\mathrm{dec},\bm{\sigma}_t^\mathrm{dec}}$, describing how likely it is that the true data point falls in the model output distribution. 
PyTorch code is available \texttt{https://github.com/dgedon/DeepSSM\_SysID}. 

\subsection{Toy Problem: Linear Gaussian System}

Consider the following linear system with process noise $\mat{v}_k\sim \Normal{0, 0.5\cdot\mat{I}}$ and measurement noise $\mat{w}_k\sim \Normal{0, 1}$ 
\begin{subequations}
\begin{align}
    \mat{x}_{k+1} &= \begin{bmatrix} 0.7 & 0.8 \\ 0 & 0.1
    \end{bmatrix} \mat{x}_k + \begin{bmatrix} -1 \\ 0.1\end{bmatrix} \mat{u}_k + \mat{v}_k, \\
    \mat{y}_k &= \begin{bmatrix} 1 & 0 \end{bmatrix} \mat{x}_k + \mat{w}_k.
\end{align}
\end{subequations}
The models are trained and validated with 2~000 samples and tested on 5\thinspace000 samples. The same number of layers in the NNs is taken for all models but with different number of neurons per layer. A grid search for the selection of the best architecture is performed with $h_\mathrm{dim}=\{50, 60, 70 ,80\}$ and $z_\mathrm{dim}=\{2,5,10\}$. Here $n_\mathrm{layer}=1$ is chosen due to the simplicity of the experiment. For all models the architecture with the lowest RMSE value is presented.

%

The deep SSMs are compared with two methods. First, a linear model is identified using SSEST from the system identification toolbox, \cite{ljungSystemIdentificationToolbox2018} with the true system order of $2$. SSEST also estimates the output variance, which is used as baseline. Second, the true system matrices as best possible linear model are run in open loop without noise.

The results are listed in Table~\ref{tab4.1:lgssm:results}; the models are listed with increasing complexity. For the deep SSMs the values are averaged over 50 identified models and for the baseline methods over 500 identifications, since these methods are computationally less expensive. The results indicate that the deep SSMs can reach an accuracy close to the state of the art methods. Note that SSEST assumes a linear model, whereas the deep SSMs fit a flexible, nonlinear model. The table also shows that a more complex deep SSM yields more accurate results. 
An open loop plot with mean and confidence interval of $\pm3$ standard deviation for the identified models by STORN and SSEST (only mean) is given in Fig.~\ref{fig4.1:lgssm:timeevolution} and compared to the ground truth. The uncertainty is captured well, but it is conservatively overestimated.

\begin{figure}[ht]
	\begin{minipage}{0.95\columnwidth}
		\captionof{table}{Results for linear Gaussian toy problem.}
		\label{tab4.1:lgssm:results}
		\begin{tabular}{l|lll}
    		Model & RMSE & NLL & ($h_\mathrm{dim}$,$z_\mathrm{dim}$) \\
    		\hline
    		VAE-RNN     & 1.56 & 1.95 & (80,10) \\
    		VRNN-Gauss-I& 1.48 & 1.82 & (50,5) \\
    		VRNN-Gauss  & 1.47 & 1.85 & (80,2) \\
    		VRNN-GMM-I  & 1.45 & 1.80 & (70,10) \\
    		VRNN-GMM    & 1.43 & 1.79 & (50,5) \\
    		STORN       & 1.43 & 1.79 & (60,5) \\
    		\hline
    		SSEST       & 1.41 & 1.78 & - \\
    		True lin. model & 1.34 & - & - \\
	    \end{tabular}
	\end{minipage}
\end{figure}

\begin{figure}[hpt]
	\centering
	\includegraphics[width=\columnwidth]{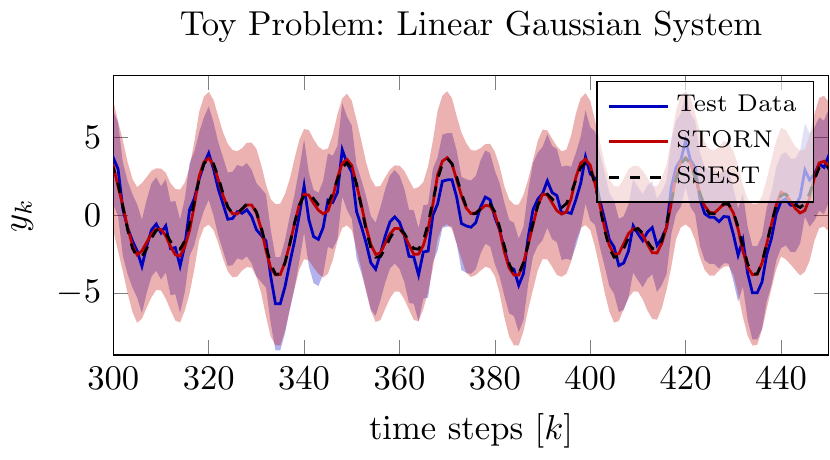}
	\caption{Toy problem: results of open loop run for test data, STORN (both with $\mu\pm~3\sigma$) and SSEST. Shaded area depicts uncertainty.}
	\label{fig4.1:lgssm:timeevolution}
\end{figure}

\subsection{Narendra-Li Benchmark}

The dynamics of the Narendra-Li benchmark are given by \cite{narendra1996nonlinearbenchmark} with additional measurement noise from \cite{stenmanModelDemandAlgorithms1999}. The benchmark is designed as a highly nonlinear but non-physical, fictional system. \ifarxiv For more details, see the appendix. \fi

This benchmark is evaluated for a varying number of training samples in $\left[2~000; 60~000\right]$. For each identification 5\thinspace000 validation samples and the same 5\thinspace000 test samples are used. A gridsearch is performed to choose architecture parameters, revealing, that in general it is advantageous to have larger networks. Hence, for comparability all models are run with $h_\mathrm{dim}=60$, $z_\mathrm{dim}=10$ and $n_\mathrm{layer}=1$. No batch normalization is applied. 

\begin{figure}[thpb]
	\centering
	\includegraphics{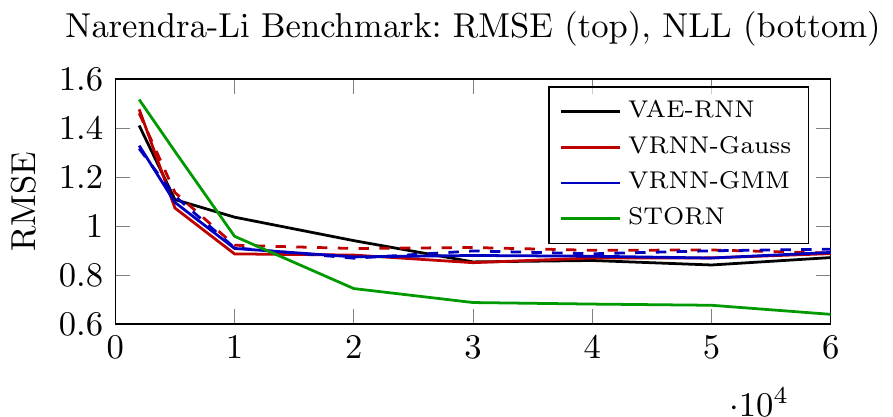}
	\includegraphics{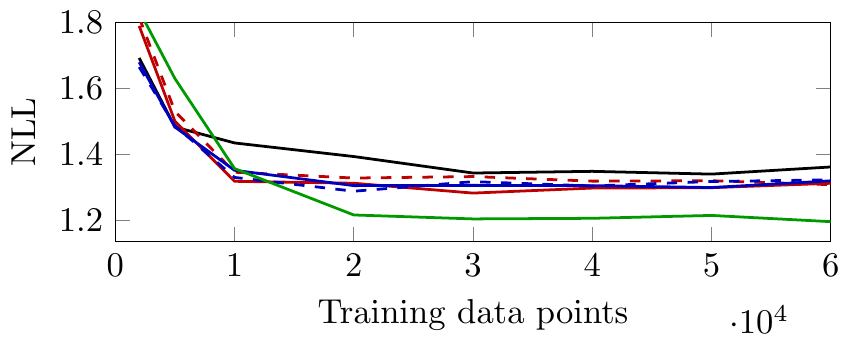}
	\caption{Narendra-Li benchmark: RMSE and NLL for varying number of training data points. VRNN-Gauss-I and VRNN-GMM-I with dashed lines.}
	\label{fig4.2:narendarli:ndata}
\end{figure}

The results are plotted in Fig.~\ref{fig4.2:narendarli:ndata} and show averaged RMSE and NLL values over 30 identified models for varying training data sizes. Generally, more training data yields more accurate estimates, both in terms of RMSE and NLL. After a specific amount of training data, the identification results stop to improve. This plateau indicates that the chosen model is saturated. Larger models could be more flexible to decrease the values even further. Specifically, the STORN model outperforms the other models, all of which show similar performance. This is due to the enhanced flexibility in STORN via the use of a second recurrent network in the inference, allowing for the learned state representations $\mat{z}_t$ to be more accurate.

The lowest RMSE values of each model are in Table~\ref{tab4.2:narendrali:results} compared with results from literature. The methods compared against do not estimate uncertainty, therefore NLL cannot be provided. Table~\ref{tab4.2:narendrali:results} also includes the required number of samples to obtain the given performance. The table indicates that deep SSMs require more samples for learning than classic models which is in line with general deep learning experience. Despite the performance gap, we believe this research to be of interest in areas where many datapoints are available and deep SSM can provide an accurate black-box  model. One reason for the performance gap can be that gray-box models from literature are compared with deep SSMs which are black-box models. The results indicate that in particular STORN reaches RMSE values close to gray-box models.

An open-loop run for identified STORN model compared with the true data is given in Fig.~\ref{fig4.2:narendarli:timeevolution}. Mean value and $\pm3$ standard deviations are shown. The figure highlights: First, the complex nonlinear dynamics are identified well. Second, the uncertainty bounds are captured but are much more conservative than the true bounds. 

\begin{figure}[ht]
	\begin{minipage}{0.95\columnwidth}
		\captionof{table}{Results for the Narendra-Li benchmark.}
	    \label{tab4.2:narendrali:results}
		\begin{tabular}{l|lll}
    		Model & RMSE & NLL & Samples \\
    		\hline
    		VAE-RNN     & 0.84 & 1.34 & 50\thinspace000\\
    		VRNN-Gauss-I& 0.89 & 1.31 & 60\thinspace000\\
    		VRNN-Gauss  & 0.85 & 1.28 & 30\thinspace000\\
    		VRNN-GMM-I  & 0.87 & 1.29 & 20\thinspace000\\
    		VRNN-GMM    & 0.87 & 1.30 & 50\thinspace000\\
    		STORN       & 0.64 & 1.20 & 60\thinspace000\\
    		\hline
    		Multivariate adaptive & 0.46 & - & 2\thinspace000\\
    		~~~~regression splines & & \\
    		Adapt. hinging hyperplanes & 0.31 & - & 2\thinspace000\\
    		Model-on-demand & 0.46 & - & 50\thinspace000\\ 
    		Direct weight optimization & 0.43 & - & 50\thinspace000 \\
    		
    		Basis function expansion & 0.06 & - & 2\thinspace000\\ 
    	\end{tabular}
	\end{minipage}
\end{figure}

\begin{figure}[htb]
	\centering
	\includegraphics{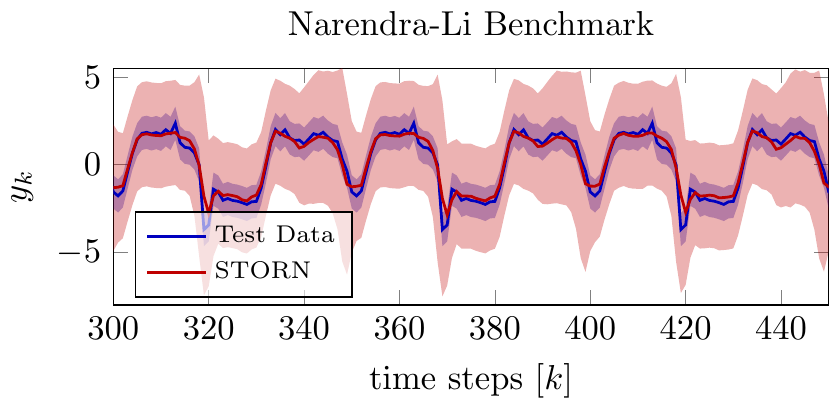}
	\caption{Narendra-Li benchmark: Time evaluation of true system and STORN with uncertainties.}
	\label{fig4.2:narendarli:timeevolution}
\end{figure}

\subsection{Wiener-Hammerstein Process Noise Benchmark} \label{sec:4c:WH}

The WH benchmark with process noise by \cite{schoukensWienerHammersteinBenchmarkProcess2016} provides measured input-output data from an electric circuit. The system can be described by a nonlinear WH model which has a nonlinearity between two linear dynamic systems. Process noise enters before the nonlinearity making the benchmark particularly difficult. 

The training data consist of 8\thinspace192 samples where the input is a faded multisine realization. The validation data are taken from the same data set but a different realization. The test data set consists of 16\thinspace384 samples, one multisine realization and one swept sine. Preliminary tests indicate that a longer training sequence length yield more accurate results, hence a length of 2\thinspace048 points is used. This benchmark is evaluated for varying sizes of the deep SSM layers. Here $h_\mathrm{dim}=\{30,40,50,60\}$ with constant $z_\mathrm{dim}=3$ and $n_\mathrm{layer}=3$.

\begin{figure}[ht] 
	\centering
	\includegraphics[width=\columnwidth]{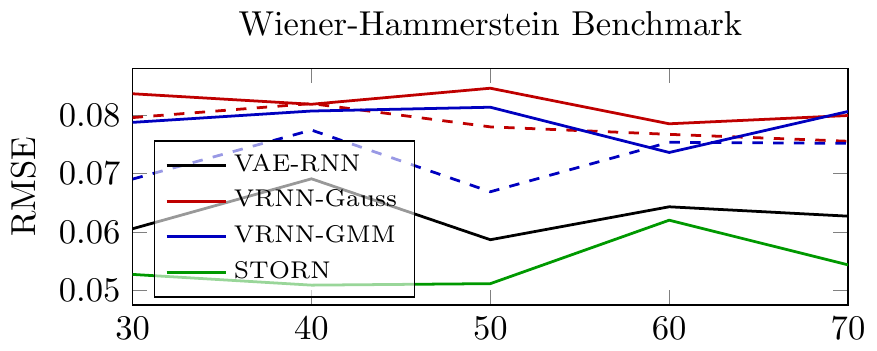}
	\includegraphics[width=\columnwidth]{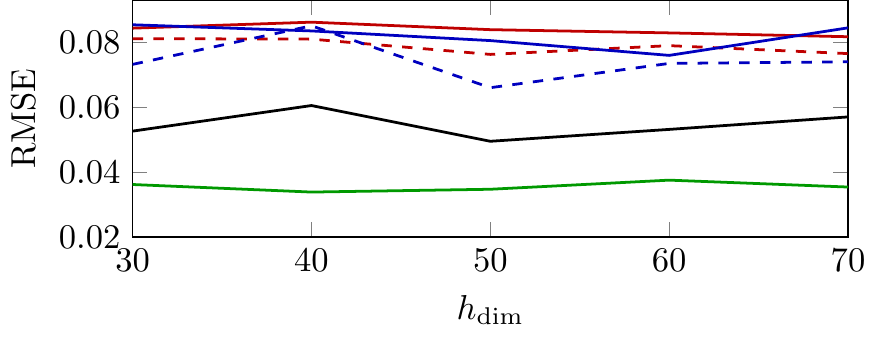}
	\caption{WH benchmark: RMSE of for mulitsine (top) and swept sine (bottom) test signal for varying $h_\mathrm{dim}$.}
	\label{fig4.3:wh:results:hvary}
\end{figure}

The resulting RMSE values for the multisine and swept sine test sequence are presented in Fig.~\ref{fig4.3:wh:results:hvary}. The lowest RMSE values of the plot are in Table~\ref{tab4.3:wh:results} compared to state of the art methods from the literature. The values are presented as averages over 20 identified models. The plot indicates that the influence of $h_\mathrm{dim}$ is rather limited. Larger values and therefore larger NNs in general tend to result in more accurate identification results. Again, STORN yields the best results, while also the very simple VAE-RNN identifies this complex benchmark well. The jagged behaviour of the plot may arise since the chosen identification data set only consists of two realizations. Therefore the randomness over the multiple experiments originates mainly from random initialization of the weights in the NNs. The difference to results from literature in Table~\ref{tab4.3:wh:results} could result because these methods are gray-box models and incorporate system knowledge.

\begin{figure}[ht]
	\begin{minipage}{0.95\columnwidth}
		\captionof{table}{Results in RMSE for WH benchmark.}
	    \label{tab4.3:wh:results}
    	\begin{tabular}{l|ccccc}
    		Model & swept sine & multisine \\
    		\hline
    		VAE-RNN     & 0.050 & 0.059 \\ 
    		VRNN-Gauss-I& 0.076 & 0.076 \\ 
    		VRNN-Gauss  & 0.082 & 0.079 \\ 
    		VRNN-GMM-I  & 0.066 & 0.067 \\ 
    		VRNN-GMM    & 0.076 & 0.074 \\ 
    		STORN       & 0.034 & 0.051 \\ 
    		\hline
    		NOBF & $\approx$0.2 & $<$0.3 \\  
    		NFIR & $<$0.05 & $<$0.05 \\  
    		NARX & $<$0.05 & $\approx$0.05 \\   
    		PNLSS & 0.022 & 0.038 \\  
    		Best Linear Approx. & - & 0.035 \\  
    		ML & - & 0.016 \\  
    		SMC & 0.014 & 0.015 \\  
    	\end{tabular}
    \end{minipage}
\end{figure}
\section{Conclusion and Future Work}

This paper provides an introduction to deep SSMs as an extension to SSMs using highly flexible NNs and elaborates the parameter learning method based on variational inference. Six deep SSMs are implemented and applied to three system identification problems to benchmark their potential. The results indicate that the class of deep SSMs is competitive to classic identification methods. 
Therefore, the toolbox of nonlinear identification methods is extended by a new model class based on deep learning. Deep SSMs also estimate the uncertainty in the dynamics by its probabilistic nature, which appears to be as conservative as established uncertainty quantification methods. This conservative behavior is in line with the existing literature on variational inference of deep learning models.

This study concerns a subclass of deep SSMs based on variational inference methods. Future work should study a broader class of deep SSMs and more nonlinear system identification benchmarks should be considered. It is of high interest to use deep SSM in automatic control like e.g. model predictive control and to elaborate how to exploit the latent state variables.

\bibliography{ifacconf}

\begin{thebibliography}{23}
\providecommand{\natexlab}[1]{#1}
\providecommand{\url}[1]{\texttt{#1}}
\providecommand{\urlprefix}{URL }
\expandafter\ifx\csname urlstyle\endcsname\relax
  \providecommand{\doi}[1]{doi:\discretionary{}{}{}#1}\else
  \providecommand{\doi}{doi:\discretionary{}{}{}\begingroup
  \urlstyle{rm}\Url}\fi

\bibitem[{Alias Parth~Goyal et~al.(2017)Alias Parth~Goyal, Sordoni,
  C{\^o}t{\'e}, Ke, and Bengio}]{aliasparthgoyalZForcingTrainingStochastic2017}
Alias Parth~Goyal, A.G., Sordoni, A., C{\^o}t{\'e}, M.A., Ke, N.R., and Bengio,
  Y. (2017).
\newblock Z-{{Forcing}}: {{Training Stochastic Recurrent Networks}}.
\newblock In \emph{Advances in {{Neural Information Processing Systems}} 30},
  6713--6723. {Curran Associates, Inc.}

\bibitem[{Andersson et~al.(2019)Andersson, Ribeiro, Tiels, Wahlstr{\"o}m, and
  Sch{\"o}n}]{anderssonDeepConvolutionalNetworks2019}
Andersson, C., Ribeiro, A.H., Tiels, K., Wahlstr{\"o}m, N., and Sch{\"o}n, T.B.
  (2019).
\newblock Deep {{Convolutional Networks}} in {{System Identification}}.
\newblock In \emph{Proceedings of the 58th IEEE Conference on Decision and
  Control}. {Nice, France}.

\bibitem[{{\AA}str{\"o}m and
  Eykhoff(1971)}]{astromSystemIdentificationSurvey1971}
{\AA}str{\"o}m, K.J. and Eykhoff, P. (1971).
\newblock System identification\textemdash{{A}} survey.
\newblock \emph{Automatica}, 7(2), 123--162.

\bibitem[{Bayer and Osendorfer(2015)}]{bayerLearningStochasticRecurrent2015}
Bayer, J. and Osendorfer, C. (2015).
\newblock Learning {{Stochastic Recurrent Networks}}.
\newblock \emph{arXiv:1411.7610}.

\bibitem[{Blei et~al.(2017)Blei, Kucukelbir, and
  McAuliffe}]{bleiVariationalInferenceReview2017}
Blei, D.M., Kucukelbir, A., and McAuliffe, J.D. (2017).
\newblock Variational {{Inference}}: {{A Review}} for {{Statisticians}}.
\newblock \emph{Journal of the American Statistical Association}, 112(518),
  859--877.

\bibitem[{Cho et~al.(2014)Cho, {van Merri{\"e}nboer}, Bahdanau, and
  Bengio}]{choPropertiesNeuralMachine2014}
Cho, K., {van Merri{\"e}nboer}, B., Bahdanau, D., and Bengio, Y. (2014).
\newblock On the {{Properties}} of {{Neural Machine Translation}}:
  {{Encoder}}\textendash{{Decoder Approaches}}.
\newblock In \emph{Proceedings of {{SSST}}-8, {{Eighth Workshop}} on
  {{Syntax}}, {{Semantics}} and {{Structure}} in {{Statistical Translation}}},
  103--111. {Association for Computational Linguistics}, {Doha, Qatar}.

\bibitem[{Chung et~al.(2015)Chung, Kastner, Dinh, Goel, Courville, and
  Bengio}]{chungRecurrentLatentVariable2015}
Chung, J., Kastner, K., Dinh, L., Goel, K., Courville, A.C., and Bengio, Y.
  (2015).
\newblock A {{Recurrent Latent Variable Model}} for {{Sequential Data}}.
\newblock In \emph{Advances in {{Neural Information Processing Systems}} 28},
  2980--2988.

\bibitem[{Fraccaro(2018)}]{fraccaroDeepLatentVariable2018}
Fraccaro, M. (2018).
\newblock \emph{Deep {{Latent Variable Models}} for {{Sequential Data}}}.
\newblock Ph.D. thesis, DTU Compute.

\bibitem[{Fraccaro et~al.(2016)Fraccaro, S{\o}nderby, Paquet, and
  Winther}]{fraccaroSequentialNeuralModels2016}
Fraccaro, M., S{\o}nderby, S.K., Paquet, U., and Winther, O. (2016).
\newblock Sequential neural models with stochastic layers.
\newblock In \emph{Proceedings of the 30th {{International Conference}} on
  {{Neural Information Processing Systems}}}, 2207--2215. {Barcelona, Spain}.

\bibitem[{Goodfellow et~al.(2016)Goodfellow, Bengio, and
  Courville}]{goodfellow2016DeepLearning}
Goodfellow, I., Bengio, Y., and Courville, A.C. (2016).
\newblock \emph{Deep {{Learning}}}.
\newblock {MIT Press}.

\bibitem[{Goodfellow et~al.(2014)Goodfellow, {Pouget-Abadie}, Mirza, Xu,
  {Warde-Farley}, Ozair, Courville, and
  Bengio}]{goodfellow2014GenerativeAdversarialNets}
Goodfellow, I., {Pouget-Abadie}, J., Mirza, M., Xu, B., {Warde-Farley}, D.,
  Ozair, S., Courville, A., and Bengio, Y. (2014).
\newblock Generative {{Adversarial Nets}}.
\newblock In \emph{Advances in {{Neural Information Processing Systems}} 27},
  2672--2680.

\bibitem[{Hochreiter and Schmidhuber(1997)}]{hochreiterLongShortTermMemory1997}
Hochreiter, S. and Schmidhuber, J. (1997).
\newblock Long {{Short}}-{{Term Memory}}.
\newblock \emph{Neural Comput.}, 9(8), 1735--1780.

\bibitem[{Kingma and Ba(2015)}]{kingmaAdamMethodStochastic2017}
Kingma, D.P. and Ba, J. (2015).
\newblock Adam: {{A Method}} for {{Stochastic Optimization}}.
\newblock In \emph{3rd {{International Conference}} on {{Learning
  Representations}}, ({{ICLR}})}. {San Diego, CA, USA}.

\bibitem[{Kingma and Welling(2014)}]{kingmaAutoEncodingVariationalBayes2013}
Kingma, D.P. and Welling, M. (2014).
\newblock Auto-{{Encoding Variational Bayes}}.
\newblock In \emph{Proceedings of the {{International Conference}} on
  {{Learning Representations}} ({{ICLR}})}. {Banff, Canada}.

\bibitem[{Liwicki and Bunke(2005)}]{liwickiIAMOnDBOnlineEnglish2005}
Liwicki, M. and Bunke, H. (2005).
\newblock {{IAM}}-{{OnDB}} - an on-line {{English}} sentence database acquired
  from handwritten text on a whiteboard.
\newblock In \emph{Eighth {{International Conference}} on {{Document Analysis}}
  and {{Recognition}} ({{ICDAR}}'05)}, 956--961 Vol. 2.

\bibitem[{Ljung(1999)}]{ljungSystemIdentificationTheory1999}
Ljung, L. (1999).
\newblock System identification: theory for the user.
\newblock \emph{PTR Prentice Hall, Upper Saddle River, NJ}.

\bibitem[{Ljung(2018)}]{ljungSystemIdentificationToolbox2018}
Ljung, L. (2018).
\newblock \emph{System Identification Toolbox: {{The Manual}}}.
\newblock {The MathWorks Inc.}, {Natick, MA, USA}, 9th edition 2018 edition.

\bibitem[{Narendra and Li(1996)}]{narendra1996nonlinearbenchmark}
Narendra, K.S. and Li, S.M. (1996).
\newblock Neural networks in control systems.
\newblock chapter~11, 347--394. Lawrence Erlbaum Associates, Hillsdale, NJ,
  USA.

\bibitem[{Prahallad et~al.(2013)Prahallad, Vadapalli, Elluru, Mantena,
  Pulugundla, Bhaskararao, Murthy, King, Karaiskos, and
  Black}]{prahalladBlizzardChallenge20132013}
Prahallad, K., Vadapalli, A., Elluru, N.K., Mantena, G.V., Pulugundla, B.,
  Bhaskararao, P., Murthy, H.A., King, S.J., Karaiskos, V., and Black, A.W.
  (2013).
\newblock The {{Blizzard Challenge}} 2013 - {{Indian Language Tasks}}.

\bibitem[{Roweis(1998)}]{roweisEMAlgorithmsPCA1998}
Roweis, S.T. (1998).
\newblock {{EM Algorithms}} for {{PCA}} and {{SPCA}}.
\newblock In \emph{Advances in {{Neural Information Processing Systems}} 10},
  626--632.

\bibitem[{Sch{\"o}n et~al.(2011)Sch{\"o}n, Wills, and
  Ninness}]{schonSystemIdentificationNonlinear2011}
Sch{\"o}n, T.B., Wills, A., and Ninness, B. (2011).
\newblock System identification of nonlinear state-space models.
\newblock \emph{Automatica}, 47(1), 39--49.

\bibitem[{Schoukens and
  Noel(2017)}]{schoukensWienerHammersteinBenchmarkProcess2016}
Schoukens, M. and Noel, J.P. (2017).
\newblock Wiener-{{Hammerstein}} benchmark with process noise.
\newblock \emph{20th IFAC World Congress}, 50(1), 448--453.

\bibitem[{Stenman(1999)}]{stenmanModelDemandAlgorithms1999}
Stenman, A. (1999).
\newblock \emph{Model on Demand: Algorithms, Analysis and Applications}.
\newblock Number 571 in Link{\"o}ping Studies in Science and Technology
  {{Dissertation}}. {Link{\"o}ping} {University}.

\end{thebibliography}

\ifarxiv
    \newpage
    \onecolumn
\appendix

\section{Graphical Models for Deep SSMs}
\label{app:A}

In section~\ref{sec2} the focus lies on the most simple deep SSM, namely the VAE-RNN. The loss function is then derived based on the graphical model for the VAE-RNN in Fig~\ref{fig2.2:vae:general}. Here the graphical models for all other studied deep SSMs are presented and shortly explained.

\subsection{VRNN}

The graphical model for the VRNN is shown in Fig.~\ref{figAPP:vrnn:general}. Note that the VRNN-Gauss and VRNN-GMM use the same network architecture. The difference lies in the output distribution, which is either Gaussian or a Gaussian mixture. In the VRNN the recurrence additionally makes use of the previous latent variable $\mat{z}_{t-1}$. In the generative network also the hidden state $\mat{h}_t$ has a direct influence on $\mat{\hat{y}}_t$. Both of these features give more flexibility for the network output distribution. The joint true posterior of the VRNN can be factorized according to the generative network in Fig.~\ref{figAPP:vrnn:general} as
\begin{align}
	p_\theta(\mat{y}_{1:T},\mat{z}_{1:T},\mat{h}_{1:T}|\mat{u}_{1:T},\mat{h}_0) &= p_\theta(\mat{y}_{1:T}|\mat{z}_{1:T},\mat{h}_{1:T}) \times  \notag\\ 
	\times p_\theta(\mat{z}_{1:T}|\mat{h}_{1:T})&\widetilde{p}(\mat{h}_{1:T}|\mat{z}_{1:T},\mat{u}_{1:T},\mat{h}_0).
\end{align}
The joint approximate posterior of the VRNN factorization according to the inference network as
\begin{align}
	q_\phi(\mat{z}_{1:T},\mat{h}_{1:T}|\mat{y}_{1:T},\mat{u}_{1:T},\mat{h}_0) &= q_\phi(\mat{z}_{1:T}|\mat{y}_{1:T},\mat{h}_{1:T})\times \notag \\
	\times& \widetilde{p}(\mat{h}_{1:T}|\mat{z}_{1:T},\mat{u}_{1:T},\mat{h}_0).
\end{align}
\begin{figure}[htb!]
	\centering
	\begin{subfigure}[t]{0.25\textwidth}
		\centering
		\includegraphics[width=0.6\textwidth]{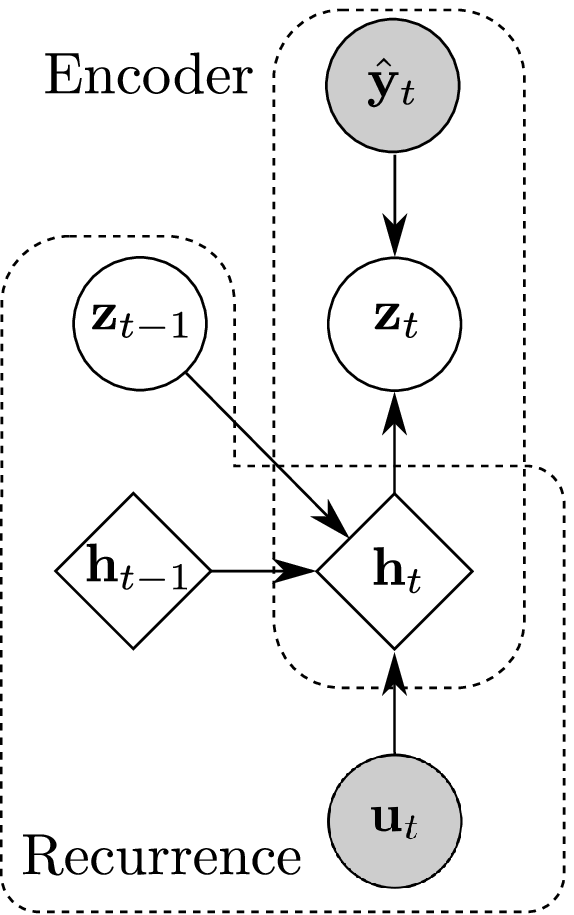}
		\caption{Inference network}
	\end{subfigure}%
	~
	\begin{subfigure}[t]{0.25\textwidth}
		\centering
		\includegraphics[width=0.7\textwidth]{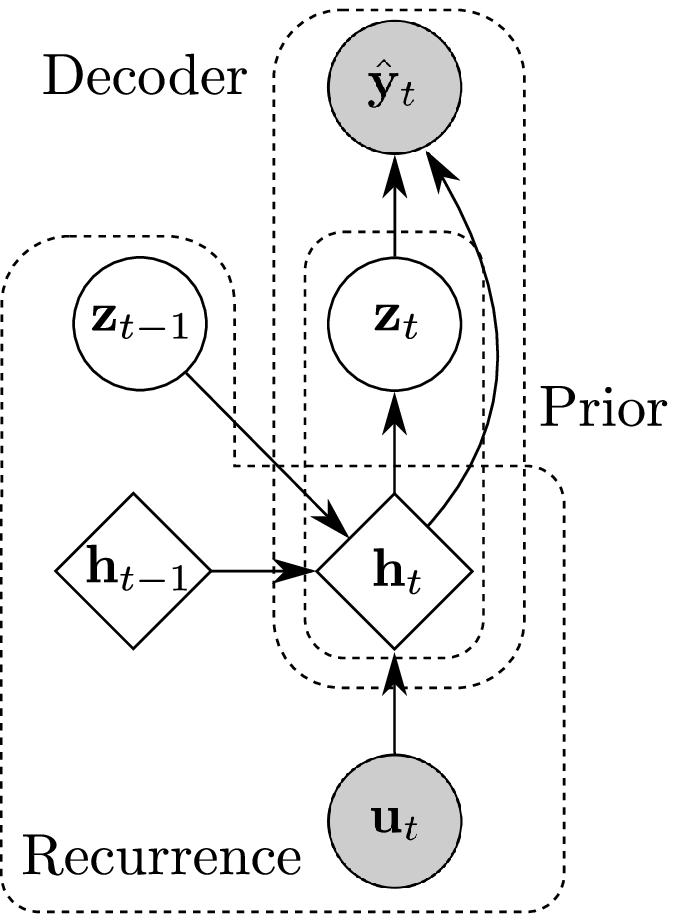}
		\caption{Generative network}
	\end{subfigure}	
	\caption{Graphical model for VRNN.}
    \label{figAPP:vrnn:general}
\end{figure}

\subsection{VRNN-I}

The VRNN-I is a simple modification of the VRNN and its graphical model is shown in Fig.~\ref{figAPP:vrnnI:general}. The difference to the VRNN is that the prior in the generative network is static and does not change temporally by the recurrent network. Similarly, here the VRNN-Gauss-I and VRNN-GMM-I only differ in the output distribution but not in the network structure. For the VRNN-I the same true and approximate joint posterior distributions as for the VRNN above apply with the difference in the true posterior that the prior is static $p_\theta(\mat{z}_{1:T}|[0,\mat{I}])$.

\begin{figure}[htb!]
	\centering
	\begin{subfigure}[t]{0.25\textwidth}
		\centering
		\includegraphics[width=0.6\textwidth]{figures/APP_VRNN_Inference.eps}
		\caption{Inference network}
	\end{subfigure}%
	~
	\begin{subfigure}[t]{0.25\textwidth}
		\centering
		\includegraphics[width=0.8\textwidth]{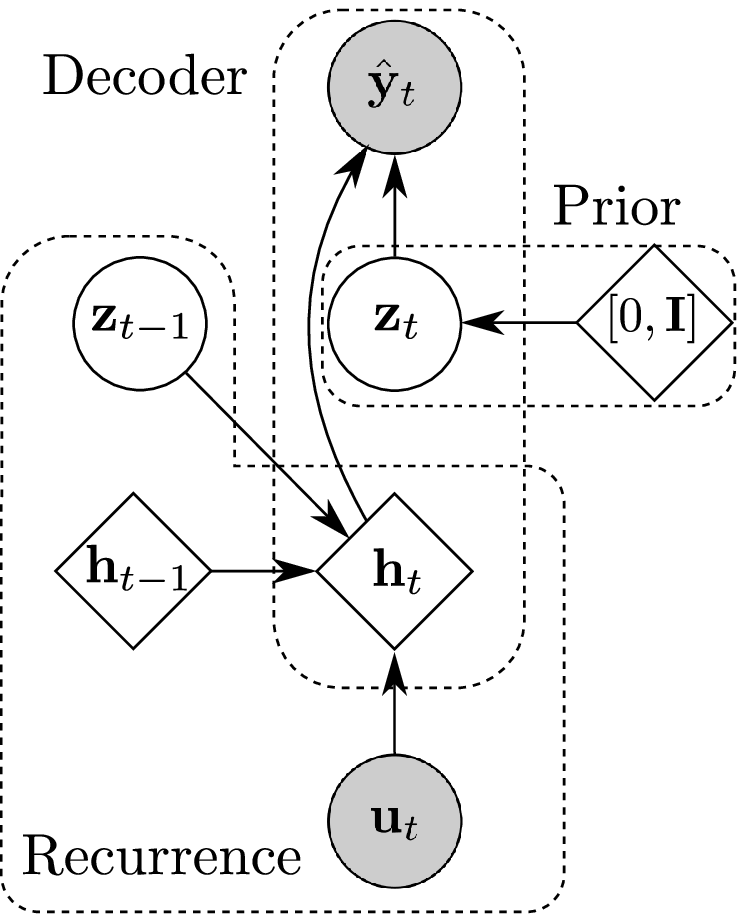}
		\caption{Generative network}
	\end{subfigure}	
	\caption{Graphical model for VRNN-I. Note that the inference network is equal to the VRNN inference network.}
    \label{figAPP:vrnnI:general}
\end{figure}

\subsection{STORN}

The graphical model for STORN is given in Fig.~\ref{figAPP:storn:general}. The main difference to the other models is the additional forward running recurrent network in the inference network. This recurrence is implemented as a GRU with the same hidden layer dimension $d_\mathrm{dim}$ as the other recurrence with $h_\mathrm{dim}$. This additional recurrence helps to encode the output distribution more precisely. Note also that in the generative network a static prior is used, similar to the VRNN-I. The joint true posterior of STORN can be factorized according to Fig.~\ref{figAPP:storn:general} as
\begin{align}
	p_\theta(\mat{y}_{1:T},\mat{z}_{1:T},\mat{h}_{1:T}|\mat{u}_{1:T},\mat{h}_0) &= p_\theta(\mat{y}_{1:T}|\mat{h}_{1:T}) \times  \notag\\ 
	\times p_\theta(\mat{z}_{1:T}|[0,\mat{I}])&\widetilde{p}(\mat{h}_{1:T}|\mat{z}_{1:T},\mat{u}_{1:T},\mat{h}_0).
\end{align}
The joint approximate posterior of STORN factorization as
\begin{align}
	q_\phi(\mat{z}_{1:T},\mat{h}_{1:T},&\mat{d}_{1:T}|\mat{y}_{1:T},\mat{u}_{1:T},\mat{h}_0,\mat{d}_0) = \notag \\
	=&~q_\phi(\mat{z}_{1:T}|\mat{d}_{1:T},\mat{h}_{1:T})\widetilde{p}(\mat{d}_{1:T}|\mat{y}_{1:T},\mat{d}_0)\times \notag\\
	&\times\widetilde{p}(\mat{h}_{1:T}|\mat{y}_{1:T},\mat{u}_{1:T},\mat{h}_0).
\end{align}

\begin{figure}[htb!]
	\centering
	\begin{subfigure}[t]{0.25\textwidth}
		\centering
		\includegraphics[width=0.7\textwidth]{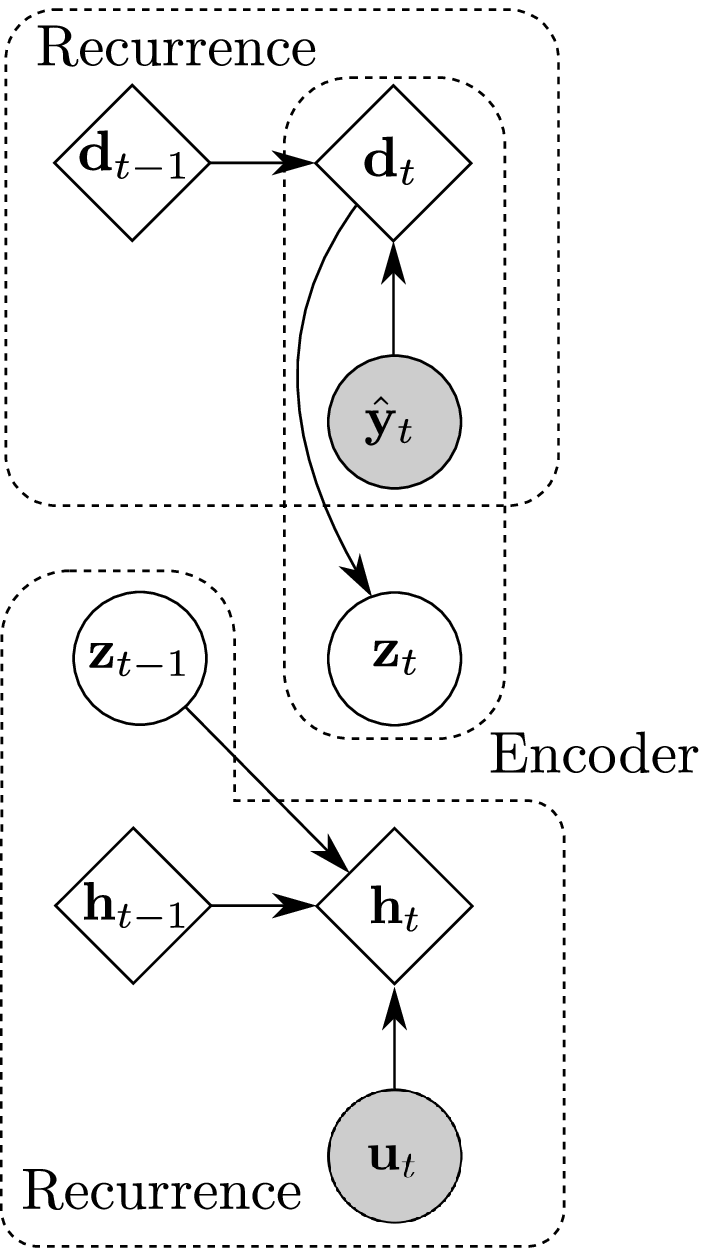}
		\caption{Inference network}
	\end{subfigure}%
	~
	\begin{subfigure}[t]{0.25\textwidth}
		\centering
		\includegraphics[width=0.75\textwidth]{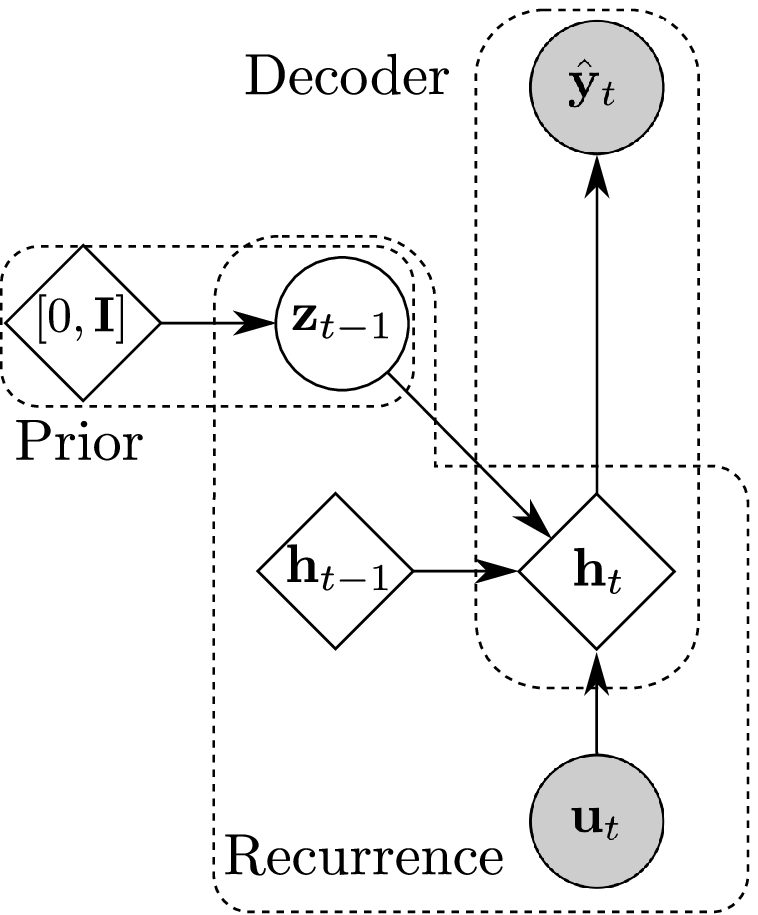}
		\caption{Generative network}
	\end{subfigure}	
	\caption{Graphical model for STORN.}
    \label{figAPP:storn:general}
\end{figure}

\section{Toy Problem: Linear Gaussian System}

For identification of the linear Gaussian toy problem an excitation input signal with uniform random noise in the range $\left[-2.5;2.5\right]$ is used for the training and validation signals. The presented results are averaged over 50 Monte Carlo identifications. For each of these identifications the training and validation sequences are drawn from a new realization with the same statistical properties. For the test data the input is given by
\begin{align}
	u_k = \sin\left(\frac{2k \pi}{10}\right) + \sin\left(\frac{2k\pi}{25}\right).
\end{align}
The same test data set is used for all identified systems in order to obtain comparable performance measures.

The numerical results from Fig.~\ref{fig4.1:lgssm:timeevolution} show that the uncertainty quantification is conservative compared to the true uncertainty bounds of the system. Here an additional figure is provided to compare state of the art uncertainty quantification as calculated by SSEST with the uncertainty quantification given by a deep SSM. In Fig.~\ref{figApp:lgssm:SSEST} this comparison is shown for the same time sequence as previously in Fig.~\ref{fig4.1:lgssm:timeevolution}. It indicates that the uncertainty quantification of STORN is comparable with the one of SSEST. Fine tuning of the hyperparameter of STORN could yield an uncertainty bound which tens towards the one of SSEST. In this experiment no fine tuning is performed.

\begin{figure}[thpb]
	\centering
	\includegraphics{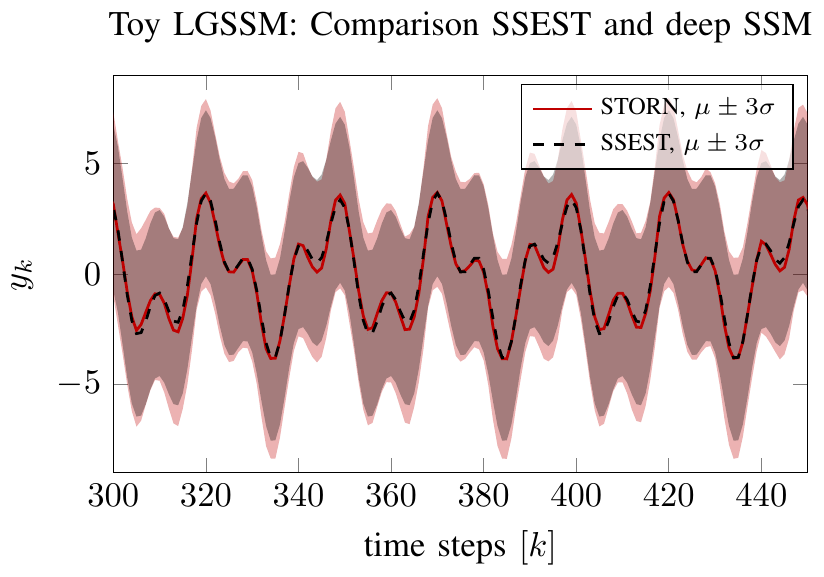}
	\caption{LGSSM toy problem: Comparison between SSEST and STORN for their uncertainty estimation.}
	\label{figApp:lgssm:SSEST}
\end{figure}

\section{Narendra-Li Benchmark}

The true dynamics of the Narendra-Li Benchmark are given by \cite{narendra1996nonlinearbenchmark} with the following second order model
\begin{align*}
	\begin{bmatrix} x_{k+1}^{(1)} \\ x_{k+1}^{(2)} \end{bmatrix} &= \begin{bmatrix} \left(\frac{x_{k}^{(1)}}{1+(x_{k}^{(1)})^2}+1\right)\sin(x_{k}^{(2)}) \\ x_{k}^{(2)} \cos(x_{k}^{(2)}) + x_{k}^{(1)} \exp(-\frac{(x_{k}^{(1)})^2+(x_{k}^{(2)})^2}{8}) + \dots \\ \dots+\frac{u_k^3}{1+u_k^2+0.5\cos(x_{k}^{(1)}+x_{k}^{(2)})}\end{bmatrix}, \\ 
	y_k &= \frac{x_{k}^{(1)}}{1+0.5\sin(x_{k}^{(2)})} + \frac{x_{k}^{(2)}}{1+0.5\sin(x_{k}^{(1)})} + e_k.
\end{align*}
Additional measurement noise is added to the original problem by \cite{stenmanModelDemandAlgorithms1999} of $e_k\sim\Normal{0,0.1}$ to make the problem more challenging. The same procedure for the excitation signals as for the linear Gaussian toy problem is used. Namely a training and validation data set where the input is uniform random noise in the range $\left[-2.5;2.5\right]$ and for the test data set the input sequence is defined by $u_k = \sin\left(\frac{2k \pi}{10}\right) + \sin\left(\frac{2k\pi}{25}\right)$. 

\section{Wiener-Hammerstein Process Noise Benchmark}

In section~\ref{sec:4c:WH} all studied deep SSMs are compared for their performance on the test data sets of the Wiener-Hammerstein process noise benchmark. Additionally, here a time evaluation is shown in Fig.~\ref{figApp:WH:Timeevoluation} for both test data sets. Note that only the first $51.2$~[ms] of the total $\approx20.97$~[ms] are shown to have well visible plots. The figure indicates an accurate identification of the complex system dynamics, which can represent the dynamics on two different test data sets. The uncertainty bounds are similarly conservative to the ones in the Narendra-Li benchmark. Tests with identifications on available data sets with more samples yield tighter uncertainty bounds but are not presented here since it would not be comparable with the comparison methods from literature.

All comparison methods in Tab.~\ref{tab4.3:wh:results} use the same amount of training samples (8192), except from PNLSS which uses 9 realization with each consisting of 8192 samples.

\begin{figure*}[thpb]
	\centering
	\includegraphics{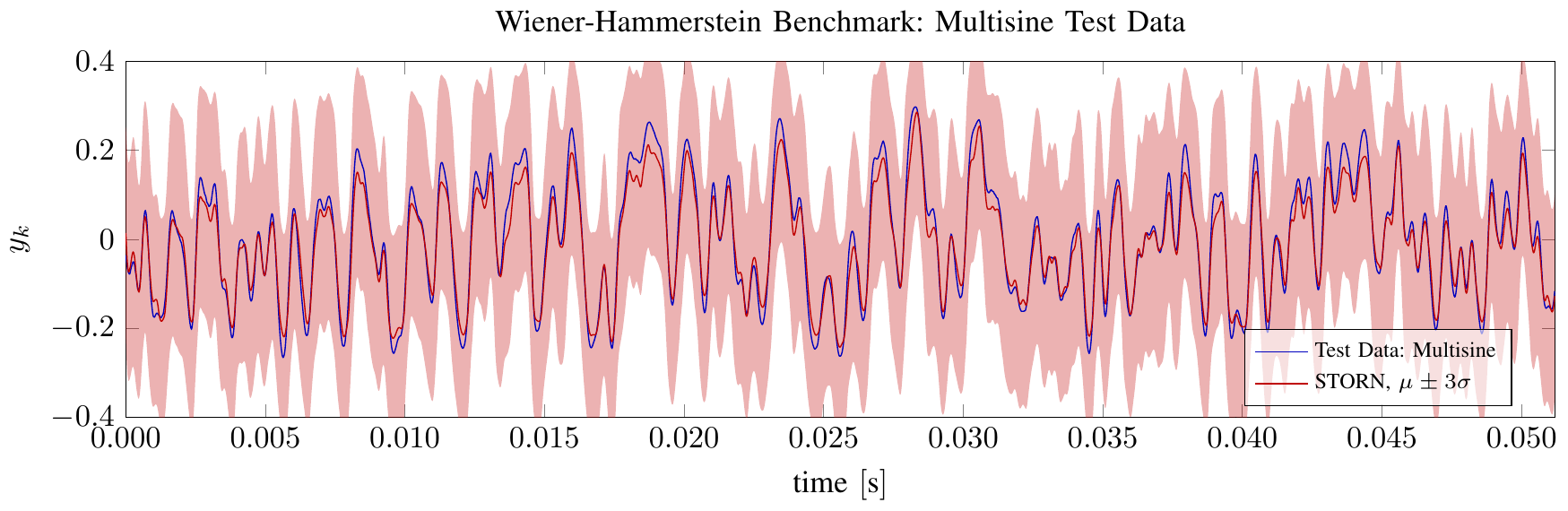}
	\includegraphics{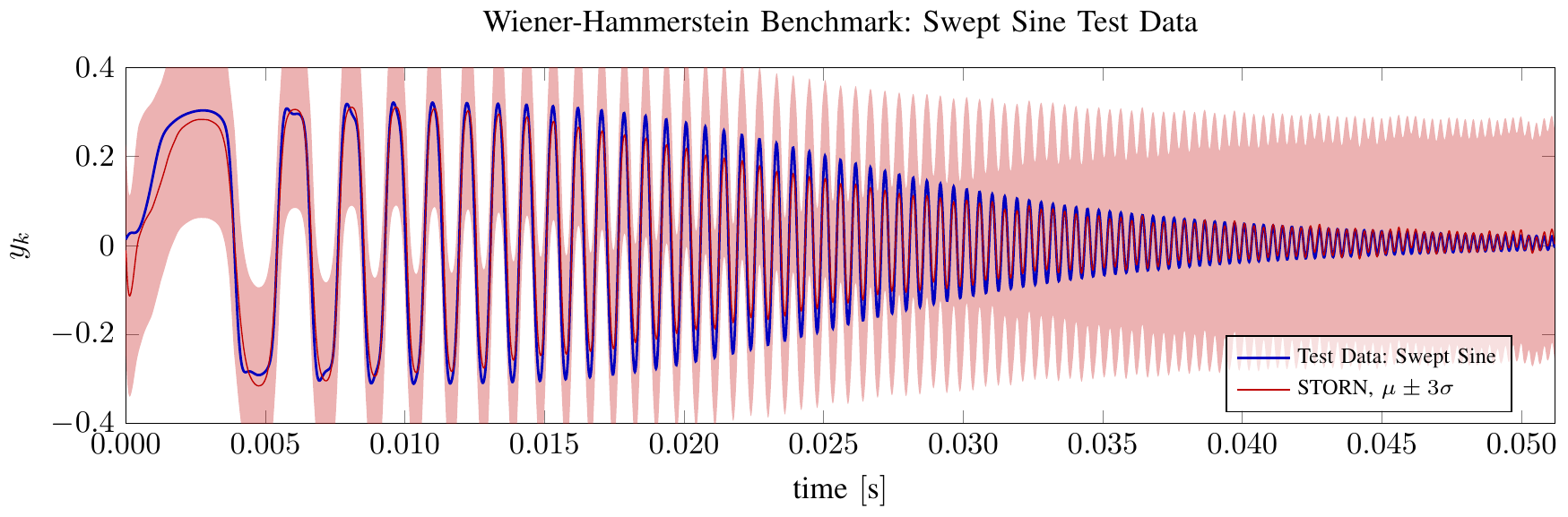}%
	\caption{Wiener Hammerstein benchmark: Time evaluation for multisine and swept sine test data set of best results from Tab.~\ref{tab4.3:wh:results}, i.e. STORN with $h_\mathrm{dim}=40$, $z_\mathrm{dim}=3$, $n_\mathrm{layers}=3$.}
	\label{figApp:WH:Timeevoluation}
\end{figure*}

\fi

\end{document}